\documentclass[12pt,preprint,natbib]{aastex}

\begin{document}

%\title{Direct Imaging Constraints on the (Non-)Existence of Protoplanets Around LkCa 15 from SCExAO/CHARIS and Keck/NIRC2}
%\title{Evidence for the Non-Existence of Protoplanets Around LkCa 15\\  from SCExAO/CHARIS and Keck/NIRC2}
\title{A Pathfinder for Imaging Extrasolar Earths from the Ground}
 % \\LkCa 15 \lowercase{bcd} are Likely (Contaminated By) Inner Disk Signals}
%\correspondingauthor{Thayne Currie}
%\email{thayne.m.currie@nasa.gov,currie@naoj.org}
\author{Thayne Currie\altaffilmark{1,2}}
\altaffiltext{1}{NASA-Ames Research Center, Moffett Blvd., Moffett Field, CA, USA}
\altaffiltext{2}{Subaru Telescope, National Astronomical Observatory of Japan, 
650 North A`oh$\bar{o}$k$\bar{u}$ Place, Hilo, HI  96720, USA}
\begin{abstract}
The Subaru Coronagraphic Extreme Adaptive Optics (SCExAO) project is an instrument on the Subaru telescope that is pushing the frontiers of what is possible with ground-based high-contrast imaging of extrasolar planets.   The system features key breakthroughs in wavefront sensing and coronagraphy to yield extremely high Strehl ratio corrections and deep planet-to-star contrasts, even for optically faint stars.   SCExAO is coupled to a near-infrared integral field spectrograph -- CHARIS -- yielding robust planet spectral characterization.   In its first full year of operations, SCExAO has already clarified the properties of candidate companions around $\kappa$ And, LkCa 15, and HD 163296, showing the former to be a likely low-gravity, planet-mass object and the latter two to be misidentified disk signals.   SCExAO's planet imaging capabilities in the near future will be further upgraded; the system is emerging as a prototype of the kind of dedicated planet-imaging system that could directly detect an Earth-like planet around a nearby low-mass star with Extremely Large Telescopes like the \textit{Thirty Meter Telescope}. 
\end{abstract}
%\keywords{} 
\section{Introduction}
Over the past 10 years, high-contrast imaging systems and now dedicated extreme adaptive optics (AO) systems on telescopes have revealed the first direct detections of young, self-luminous Jovian extrasolar planets on Solar System-like scales, up to one million times fainter than the stars they orbit \citep[e.g.][]{Marois2008}.  A key goal of the upcoming 30m-class Extremely Large Telescopes (ELTs) is to image a rocky, Earth-sized planet in the habitable zone around a nearby M star \citep{Skidmore2015}.   However, doing so requires achieving planet-to-star contrasts 100 times deeper ($\sim$ 10$^{-8}$) at 10 times closer angular separations ($\theta$ $\sim$ 0\farcs{}05--0\farcs{}1) than currently demonstrated with leading instruments like the Gemini Planet Imager (GPI) or SPHERE at the Very Large Telescope.  

\section{SCExAO Design and Performance}
The Subaru Coronagraphic Extreme Adaptive Optics project (SCExAO) utilizes and will further mature key advances in wavefront control and coronagraphy needed to move beyond recent exoplanet direct imaging capabilities to those capable of imaging an Earth with ELTs \citep{Jovanovic2015}.   The system takes as input and then sharpens a partially corrected wavefront from Subaru's facility AO system (A0-188).   Subaru's 2,000-actuator deformable mirror is driven by a modulated Pyramid wavefront sensor (WFS) \citep{Lozi2018}.   Owing to these Pyramid optics and an advanced camera used for wavefront sensing (OCAM-2K), SCExAO can target fainter stars than possible with the spatially filtered Shack-Hartmann WFSs used by GPI and SPHERE and its wavefront control loop can run faster (in practice at 2 kHz, in principle up to 3.5 kHz).   The host star's halo is further suppressed by a suite of versatile coronagraphs (e.g. a vector-vortex, shaped-pupil, and standard Lyot coronagraph).  

Starlight is fed into different optical/near-infrared (IR) science instruments for exoplanet imaging this is primarily CHARIS, a near-IR integral field spectrograph \citep{Groff2015}.   CHARIS operates in two modes.   Its low resolution (R $\sim$ 20)  ``broadband mode" covering the three major near-IR bandpasses (J, H, and K)  is well-suited for exoplanet discovery and coarse spectral characterization.   A higher-resolution (R $\sim$ 70) mode in J, H, or K allows for more detailed, follow-up characterization.

Presently, SCExAO delivers high-quality AO corrections -- $\sim$ 90\% Strehl at 1.6 µm under good conditions -- and planet-to-star contrasts ($\sim$ 10$^{-6}$ at 0\farcs{}5) similar to those from GPI and SPHERE and far superior to those achieved with facility AO systems (Figure 1).   It has reached extreme-AO-like contrasts on stars as faint as 12th magnitude in the optical.   These deep contrasts translate into new exoplanet discovery space and higher-fidelity spectra for known directly imaged exoplanets.

\begin{figure}[ht]
\centering
\includegraphics{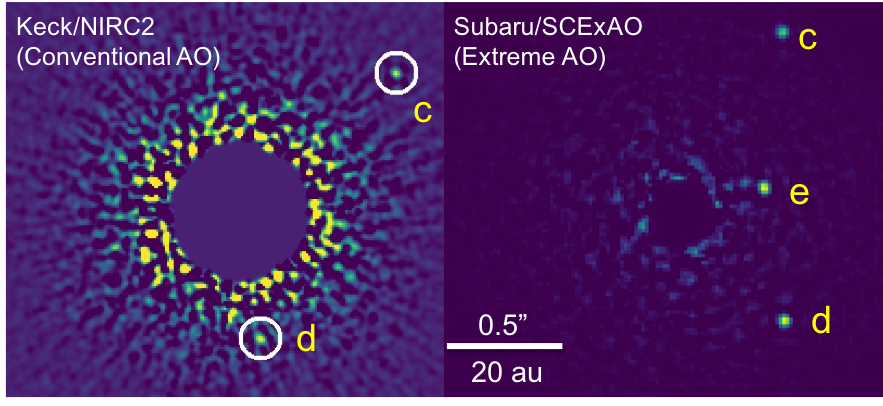}
\vspace{-0.2in}
\caption{Two near-infrared images of the HR 8799 system. The left image is from a leading conventional (facility) AO system on the Keck telescope; on the right is a SCExAO/CHARIS image.  SCExAO/CHARIS produces higher signal-to-noise detections of HR 8799 cd and a detection of HR 8799 e.}
\end{figure}

\section{SCExAO Science Results}
The first year of full science operations for SCExAO and CHARIS has yielded new insights about massive exoplanets and planet candidates.   Most notably, it provided the first high-quality JHK spectrum of $\kappa$ And b, an object whose nature -- young exoplanet or intermediate-aged brown dwarf -- has been the subject of considerable controversy.   The SCExAO/CHARIS spectrum for $\kappa$ And b showed evidence for a low surface gravity and similarity to young objects in the planet-mass regime \citep{Currie2018}.  

On the other hand, SCExAO/CHARIS data showed that candidate protoplanets around LkCa 15 and HD 163296, thought to be located within cleared regions of protoplanetary disks, are instead likely misidentified disk signals \citep{Rich2019,Currie2019}.   Even with current capabilities, SCExAO/CHARIS imaging of many nearby young stars as a part of an upcoming intensive survey should be able to discover Jovian planets on Solar System scales (5--30 au).   

\section{SCExAO: Future Upgrades and the Path to Imaging Earth-like Planets with ELTs}
Over the next 5 years, SCExAO will operate both as an exoplanet discovery/characterization engine and also as a key technology pathfinder for high-contrast imaging with ELTs.    A critical advance will be the maturation of new wavefront sensing and control techniques such as predictive control and sensor fusion \citep{MalesGuyon2018}.    Ultra-low-noise detectors -- i.e. ``Microwave Kinetic Inductance Detectors"  \citep[MKIDS;][]{Mazin2012} can drive focal-plane wavefront sensing/control (FPWFS/C) at high speed and accuracy, yielding a factor of 10--100 gain in planet-to-star contrasts, sufficient to image some exoplanets in reflected light.   The general architecture of high-contrast imaging with SCExAO -- a two-stage AO system with multiple, complementary loops for focal-plane wavefront sensing -- resembles the general architecture envisioned for extreme AO systems on some ELTs \citep[e.g. the Planetary Systems Imager on the Thirty Meter Telescope;][]{Guyon2018}.    

With SCExAO and other upcoming systems being developed over the next decade as proving grounds (e.g. Magellan/MagAO-X, Keck/KPIC, Gemini/GPI-2.0), ELTs may be capable of directly detecting and characterizing the atmospheres of rocky, habitable zone planets around the nearest M stars within the next 15-20 years.

{}

\end{document}